\documentclass[a4paper,11pt]{article}
\usepackage{pos}
\usepackage{graphicx,amsmath,color}
\usepackage{relsize}
\usepackage{enumerate}
\usepackage{slashed}
\usepackage{multirow}
\usepackage[normalem]{ulem}

\def\beq{\begin{equation}}
\def\eeq{\end{equation}}
\def\bea{\begin{eqnarray}}
\def\eea{\end{eqnarray}}

\def\bit{\begin{itemize}}
\def\eit{\end{itemize}}

\def\l{\left}
\def\r{\right}

\def\baa{\begin{array}}
\def\eaa{\end{array}}

\def\sl#1{\mathord{\not\mathrel{{\mathrel{#1}}}}}
\def\d{\partial}
\def\simgt{\mathrel{\lower2.5pt\vbox{\lineskip=0pt\baselineskip=0pt
           \hbox{$>$}\hbox{$\sim$}}}}
\def\simlt{\mathrel{\lower2.5pt\vbox{\lineskip=0pt\baselineskip=0pt
           \hbox{$<$}\hbox{$\sim$}}}}
\newcommand{\vev}[1]{ \langle {#1} \rangle }

\def\bfc{\begin{figure}\begin{center}}
\def\efc{\end{center}\end{figure}}
\def\nn{\nonumber\\}

\definecolor{chromeyellow}{rgb}{1.0, 0.65, 0.0}
\definecolor{darkcoral}{rgb}{0.8, 0.36, 0.27}
\definecolor{cadmiumgreen}{rgb}{0.0, 0.42, 0.24}

\title{Baryogenesis and leptogenesis with relativistic bubble walls}

\author*[a]{Miguel Vanvlasselaer}

\affiliation[a]{Theoretische Natuurkunde and IIHE/ELEM, Vrije Universiteit Brussel,
\& The International Solvay Institutes, Pleinlaan 2, B-1050 Brussels, Belgium}

\emailAdd{*miguel.vanvlasselaer@vub.be}

\abstract{In this talk, we study the impact of first order phase transitions with fast bubble walls on mechanisms of leptogenesis and baryogenesis. We begin our exploration with the usual leptogenesis where the breaking of $B-L$ occurs via a PT with fast walls. Then we move to a more exotic case where the $B-L$ breaking phase transition creates heavy particles in the plasma and catalizes the leptogenesis. Finally, we apply the same production mechanism to the EWPT at low energy and build a new model of EWBG. Those models are all original and contain crucial new phenomenological aspects like the emission of large amount of Gravitational waves.}

\FullConference{
 Corfu Summer institute 2023: "Workshop on Theoretical Particle Cosmology in the Early and Late Universe"\\
  APRIL 30 - MAY 6, 2023\\
  Corfu, Greece.
}

\begin{document}
\maketitle

\section{Introduction}

 One of the greatest puzzles of the early universe cosmology is the origin of  the observed excess of matter over anti-matter, which is commonly parameterized by
 \bea
 Y_{ B} \equiv \frac{n_B -n_{\bar{B}}}{s}\bigg|_0 = (8.75 \pm 0.23)\times 10^{-11}
 \eea
 with $n_B, n_{\bar{B}}$ and $s$ respectively the number density of baryons, anti-baryons and the entropy density. The second equality comes from Planck data and evolution models of the early universe\cite{Ade:2015xua}. Within the inflationary paradigm, this asymmetry calls for an explanation in terms of early universe dynamics, a dynamics which is called \emph{baryogenesis}. For a successful baryogenesis scenario,  the 
well-known Sakharov requirements should be 
satisfied~\cite{Sakharov:1967dj},
namely the 
violation of the baryon number, violation of 
$C$ and $CP$ symmetries, and the presence of an out-of-equilibrium process. Many different mechanisms can fulfil those requirements, see
\cite{Riotto:1998bt, Bodeker:2020ghk} for extensive reviews. On the top of this, baryogenesis mechanisms can be broadly classified into two different categories, depending of the SM sectors where the asymmetry originally forms: in the baryon sector, in the case of the original \emph{baryogenesis scenario} or in the lepton sector, in the case of the so-called \emph{leptogenesis scenario}\cite{FUKUGITA198645}.

First order phase transitions are very efficient ways to fulfil the out-of-equilibrium criterion, and it is used namely in the case of \emph{electroweak baryogenesis}
 \cite{Kuzmin:1985mm, Shaposhnikov:1986jp}. Several BSM models could make the EWPT first order\cite{Nelson:1991ab,Carena:1996wj,Cline:2017jvp,Bruggisser:2018mrt,Bruggisser:2018mus} (see \cite{Morrissey:2012db} for review). However, relativistic bubble walls were believed to \emph{suppress} the final baryon asymmetry\cite{Caprini:2011uz,Cline:2020jre,Dorsch:2021ubz, Shuve:2017jgj}.

In this paper we propose to reconsider this belief by discussing situations in which ultra-relativistic bubble walls actually can enhance the baryon/lepton asymmetry. We will discuss three different types of models taking advantage of fast walls: in section \ref{sec:bubble_assisted}, we study the usual leptogenesis catalized by a phase transition (see also \cite{Baldes:2021vyz} for another viable model). The Right-Handed Neutrinos (RHN) abruptly receive a large mass upon crossing the bubble wall, and decay all together, suppressing the wash-outs. 

In section \ref{sec:leptogenesis_via_production}, we study another model of leptogenesis catalized via the production of heavy states, which we identify with Majorana neutrinos. The idea is based on the  observation in\cite{Vanvlasselaer:2020niz}
that an ultra-relativistic bubble wall with Lorentz 
factor $\gamma_w \gg 1$, can produce in the plasma particles with mass 
up to $M \lesssim \sqrt{\gamma_w 
T_{\text{nuc}} \times v}$, where $T_{\text{nuc}}$ and $v$  are 
the nucleation temperature of FOPT and the scale of the 
symmetry breaking respectively. Beside being an out-of-equilibrium production channel, we will also show that this production mechanism can be naturally CP-violating. We confirm the statements above by analyzing the  CP-violating effects in the interference of tree and one loop level processes. 

 Finally, in section \ref{sec:model2}, we study a model for which the Electroweak phase transition, again with fast bubbel walls, produces heavy states and catalizes EWBG. 

 One of the interesting feature of  the class of models we discuss in this paper is that it requires ultra-relativistic bubble wall velocities and strong phase transition, and then
is generically accompanied with 
strong gravitational waves signal.

\section{When do the walls become fast? (and how fast ?)}
\label{sec:fast_walls}
For strong first order phase transitions, with moderate to strong supercooling, which will be the natural territory of exploration of this paper, the regime of expansion of the bubble walls can be determined by the balance between the driving force and the plasma pressure
\bea 
\Delta V = \Delta \mathcal{P}(\gamma_w = \gamma_w^{\rm ter}) \, , 
\eea 
where $\gamma_w^{\rm ter}$ is the terminal velocity of the bubble. If this equality is never fulfilled, we expect the wall to runaway and to keep accelerating until the bubble collisions(see however \cite{Ai:2023see,Ai:2024shx}).

In general, $\Delta \mathcal{P}(\gamma_w)$ is expected to be a very complicated function of the velocity $\gamma_w$, however, it simplifies in the regime of large $\gamma_w \gg 1$ to a sum of few contributions.
The leading order pressure due to particles gaining a mass is given by \cite{Bodeker:2009qy}
\bea 
\mathcal{P}_{\rm BM} \equiv \frac{1}{48} \sum_i g_i n_i \Delta M_i^2 T^2,
\label{eq:BPpres}
\eea 
where particle $i$ has $g_i$ degrees of freedom, and $n_i=1(2)$ for fermions (bosons), while emission of soft gauge bosons  leads to a $\gamma_w v T^3 \log (v/T)$ term\cite{Bodeker:2017cim, Gouttenoire:2021kjv, Azatov:2023xem}. Using the results presented in \cite{Gouttenoire:2021kjv, Azatov:2023xem}, we can obtain the terminal velocity for large supercoolings:
\bea 
\gamma^{\rm ter }\sim 6 \times \l(\frac{\Delta V-\Delta\mathcal{P}_{\rm LO}^{ SM}}{(100 {\rm~ GeV})^4}\r)\l(\frac{100 {\rm GeV}}{T_{\rm nuc}}\r)^3\frac{1}{\log \frac{M_z}{g T}}.
\label{eq:terminal_velo}
\eea

\section{Bubble-assisted leptogenesis}
\label{sec:bubble_assisted}
We start our exploration of the effect of fast walls on lepton and baryon yields with the simplest leptogenesis-like case\cite{Chun:2023ezg}. 
The setup we consider is as follows: The RHNs receive a large Majorana mass via the spontaneous symmetry breaking of $B-L$, which is first order and proceeds via the nucleation of bubbles (see Ref.\,\cite{Shuve:2017jgj} for the study of the second-order case).
The relevant part of the Lagrangian can be written in the mass basis of the RHNs as
\begin{align}
\label{Eq:L_RHN}
\mathcal{L}_{\text{int}} = &\frac{1}{2} \sum_{I} 
y_I \Phi \bar{N}_I^c N_I + \sum_{\alpha,\, I} Y_{D,\alpha I} H\bar{L}_{\alpha}N_I +  h.c.,
\end{align}
where $L_\alpha$ are the SM lepton doublets, $N_I$ are the three 
families of heavy right-handed neutrinos, $Y_{D,\alpha I}$ are the \emph{Dirac} Yukawa couplings between $N_I$ and $L_\alpha$, and $y_I$ are \emph{Majorana} Yukawa couplings. After the phase transition, $\langle \Phi \rangle \equiv v_\phi/\sqrt{2}$, and the type-I seesaw Lagrangian is recovered with $M_I = \frac{1}{\sqrt{2}}y_I v_\phi$.

\begin{figure}
      \centering
      \includegraphics[width=0.95\textwidth]{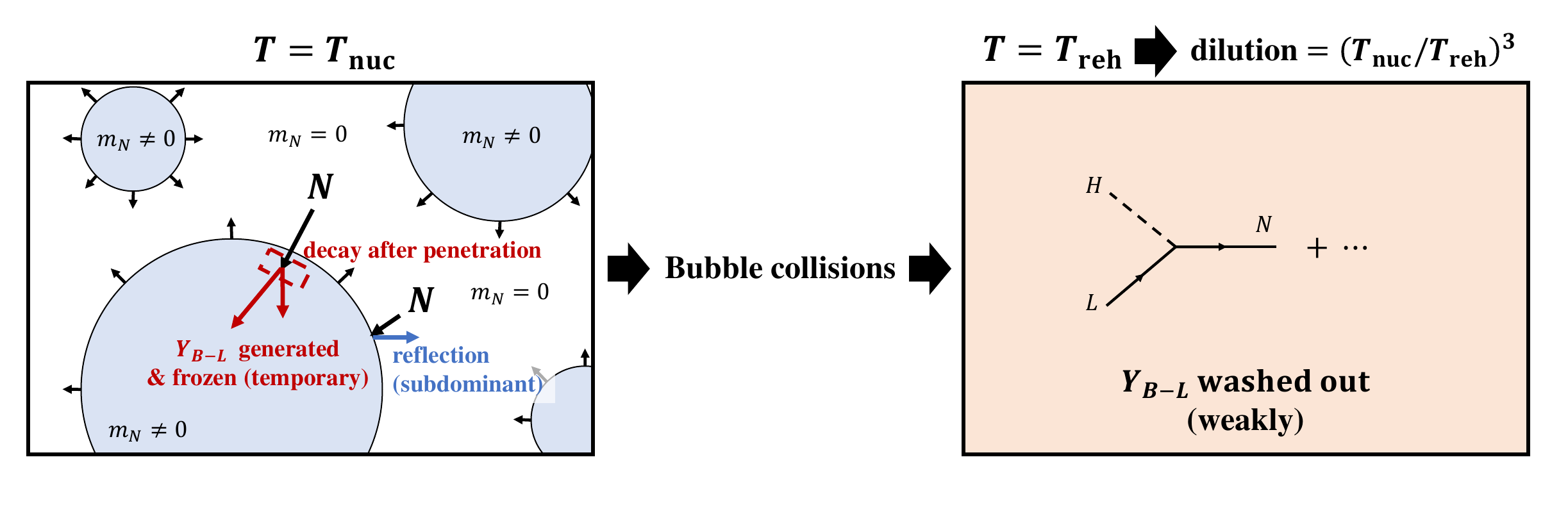}
      \caption{ Schematic picture of the bubble-assisted leptogenesis scenario during bubble expansion (left) and after bubble collisions (right). 
      }
      \label{fig:schematic}
\end{figure}

In the thermal leptogenesis scenario, the expansion of the universe fulfils the out-of-equilibrium Sakharov criterion. It is however typically a quite weak departure from equilibrium and the efficiency of traditional thermal leptogenesis is suppressed by the fact that most of heavy neutrinos decay \emph{in equilibrium}. This suppression is usually parameterized by a $\kappa_{\rm wash} \sim 0.01$
which gives the final baryon asymmetry
\bea
Y_B = Y^{\rm eq}_{N}\, \epsilon_{\rm CP}\, \kappa_{\rm sph}\, \kappa_{\rm wash} \, .
\label{eq:thermalleptorelation}
\eea
where $Y^{\rm eq}_{N}$ is the initial abundance of RHNs, $\epsilon_{\rm CP} \lesssim 0.1$ is the parameter controlling the CP violation during decay and the factor $\kappa_{\rm sph} \sim 1/3$ designates the conversion from the lepton asymmetry into a baryon asymmetry.

In the case of \emph{bubble-assisted} leptogenesis\cite{Chun:2023ezg}, the RHNs are massless outside the FOPT bubbles and become suddenly massive when the wall hits them. a cartoon of the situation is provided in Fig.\ref{fig:schematic}. The expansion of the bubble wall actually offers a new source of departure from equilibrium and opens the possibility to drastically enhance $\kappa_{\rm wash} \to 1$. This is because if the RHNs become suddenly massive, they \emph{all} decay out-of-equilibrium, not only a $\mathcal{O}(0.01)$ fraction of it (when the inverse decays are alreade decoupled). 

However, new possible sources of suppression come in compensation: i) first the RHNs might fail to enter the bubble and be reflected away, this is encapsulated by $\kappa_{\rm pen}$, ii) our setup opens new channels like $N_IN_I \to \phi \phi$, where the number density of RHN is depleted before its decay, therefore suppressing the final asymmetry and that we designates by $\kappa_{\rm dep}$, iii) at the end of the PT, entropy is injected in the plasma and dilutes any previous abundances by a factor $\left( T_{\rm nuc}/T_{\rm reh}
\right)^3$. 
At the end, the asymmetry is parameterized by 
\begin{equation}
Y_B = Y^{\rm eq}_N
\epsilon_{\rm CP} \,
\kappa_{\rm sph} \,
\kappa_{\rm pen} \,
\kappa_{\rm dep} \,
\kappa_{\rm wash} \,
\left( \frac{T_{\rm nuc}}{T_{\rm reh}}
\right)^3,
\label{eq:finalasymparam}
\end{equation}
where $Y^{\rm eq}_N$ is population of RHNs outside the bubbles. $T_{\rm reh}$ is the temperature after the FOPT and can be estimated via 
\bea
\left( \frac{T_{\rm nuc}}{T_{\rm reh}}  \right)^3
\simeq (1+\alpha_n)^{-3/4}.
\eea

Our goal is to determine the values of each parameter $Y^{\rm eq}_N ,
\epsilon_{\rm CP} ,
\kappa_{\rm sph} ,
\kappa_{\rm pen} ,
\kappa_{\rm dep} ,
\kappa_{\rm wash} ,
\left( \frac{T_{\rm nuc}}{T_{\rm reh}}
\right)^3$ as a function of the PT parameters $\alpha_n$ and $\beta$. The $\kappa_{\rm pen}$ parameter can be obtained by computing the terminal velocity of the bubble wall, via the methods discussed in Section \ref{sec:fast_walls} and using the fact that a RHN will enter the wall if its momentum all the direction of the wall expansion (in the wall frame) $p^{\rm wf}_z$ is larger than its mass inside the wall $M_I$: $p^{\rm wf}_z > M_I$. Integrating over the density of incoming RHN impinging the wall gives the fraction of entering RHN. 

Computing $\kappa_{\rm dep}$ and $
\kappa_{\rm wash}$ requires to solve the relevant Boltzmann equations. We first assume that the RHN are in kinetic equilibrium with the SM thermal bath \emph{also inside the bubble}, due to the efficient interaction $\phi N \to \phi N$ via $N$ mediation. This allows to integrate the Boltzmann equations over the momenta.  The Boltzmann equation in this procedure become
\begin{align}
\dot{n}_{N_I}+3H n_{N_I} &= 
- \sum_{A,B} \left(
        2 \langle \sigma v \rangle_{N_I N_I \to AB} \, n_{N_I}^2
        - 2 \langle \sigma v \rangle_{AB \to N_I N_I} \, n_{A} n_B
    \right)
- \Gamma_D(N_I) n_{N_I},
\label{eq:Boltzmann_nNI}
\\
\dot{n}_{B-L} +3H n_{B-L}&= 
- \sum_I \epsilon_I \Gamma_{D}(N_I)  n_{N_I} +\text{(wash-out)},
\label{eq:Boltzmann_nB-L}
\end{align}
where $\Gamma_D(N_I)$ is the total decay rate of $N_I$ and 
$\epsilon_I$ is the CP-violating parameter for $N_I$ defined by
\bea
\epsilon_{I}
\equiv
\frac
{\Gamma(N_I \to H L)
-\Gamma(N_I \to \bar H \bar L)}
{\Gamma(N_I \to H L)
+\Gamma(N_I \to \bar H \bar L)},
\eea
where $\bar L$ denotes the anti-particle of $L$.
The initial population of $N_I$ within the bubbles will be given by its massless equilibrium distribution scaled by a factor of $\kappa_{\rm pen}$:
\bea
n_{N_I}^{(0)} = 
\kappa_{\rm pen} \,
\frac{2\cdot \frac{3}{4}\cdot \zeta(3)}{\pi^2}T_{\rm nuc}^3.
\label{eq:kappa-n}
\eea

We have also checked that that the RHNs decay before the onset of bubble collisions. the depletion can decrease the initial abundance of $N_I$ via $N_I N_I \to \phi \phi$, or directly $N_I N_I \to f \bar f$, with $f$ being a SM fermion, in the case in which $B-L$ is gauged.
Finally, the flavor changing interactions $N_I N_I \leftrightarrow N_J N_J$ are also efficient and maintain equilibrium among the RHN flavors. 

This allows to define $n_N \equiv \sum_I n_{N_I}\simeq 3n_{N_1} \simeq 3n_{N_2}\simeq 3n_{N_3}$ and $Y_N \equiv n_N/s$, we obtain the familiar looking equations
\begin{align}
\label{eq:Boltzmann_YN}
z H s \, Y_N'(z)
&=
-\bar \gamma_D \l( \frac{Y_N}{Y_N^{\rm (eq)}}-1 \r)
-2 
 \gamma_{NN\to \phi\phi}
   \l(  Y_N^2- \l( Y_N^{\rm (eq)}\r)^2 \r)
+ \text{(model-dependent)}, 
\\
\label{eq:Boltzmann_YB-L}
z H s \, Y_{B-L}'(z)
&=
-\epsilon_{\rm CP} \bar \gamma_D \l( \frac{Y_N}{Y_N^{\rm (eq)}} -1 \r)
- \frac{1}{2}(c_L+ c_H)\, \bar \gamma_D \frac{Y_{B-L}}{Y^{\rm (eq)}} ,
\end{align}
where $z\equiv M_N/T$, $Y_N^{\rm (eq)} = n_N^{\rm (eq)}/(\frac{2\pi^2}{45}g_* T^3)$ 
with the equilibrium number density $n_N^{\rm (eq)}$, 
$Y_N^{\rm (eq)} = (\frac{2}{\pi^2}T^3)/(\frac{2\pi^2}{45}g_* T^3)$, and
\bea
\bar \gamma_D &\equiv& 
\sum_I \gamma_{D}(N_I)
= \sum_I n_{N_I}^{\rm (eq)} \frac{K_I(z)}{K_2(z)} \Gamma_D(N_I), 
\\
\epsilon_{\rm CP} \, \bar \gamma_D &\equiv& 
 \sum_I \epsilon_I \gamma_D(N_I),
\\
 \gamma_{NN\to \phi\phi}
&\equiv& \frac{1}{9} s^2 
\sum \langle \sigma v \rangle_{N_I N_I \to \phi \phi}.
\eea

\begin{figure}[t]
      \centering
      \includegraphics[width=0.45\textwidth]{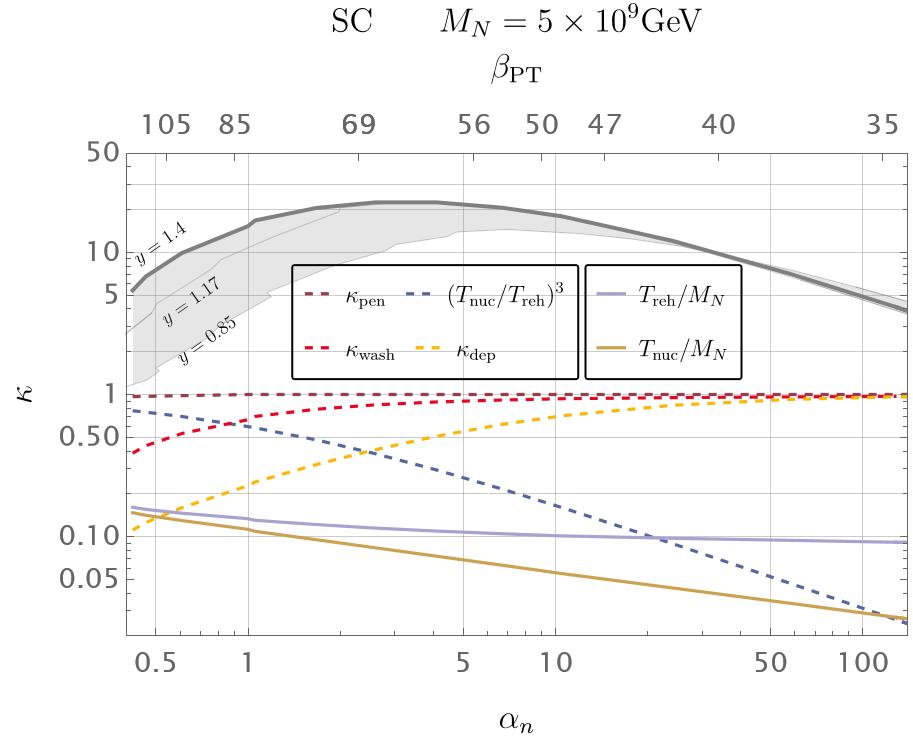}
      \includegraphics[width=0.45\textwidth]{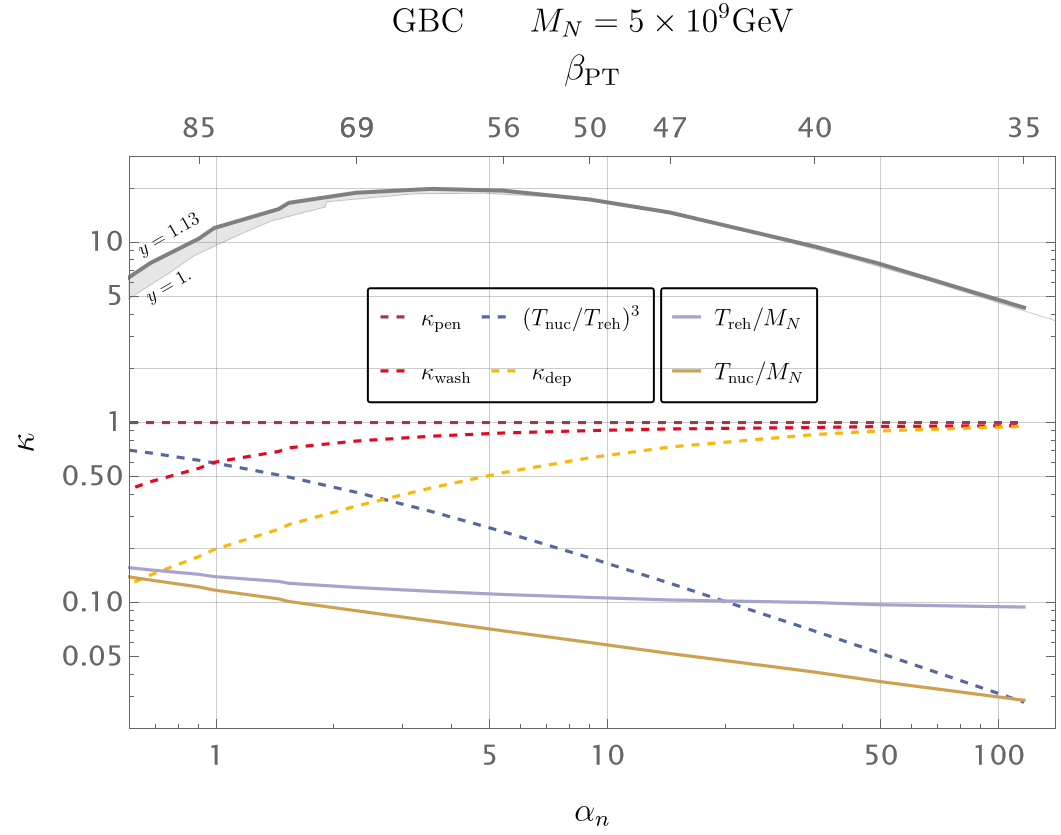}
      \caption{Comparison of the thermal leptogenesis and the bubble-assisted leptogenesis when the PT is catalized by a singlet (Left) or gauge bosons (Right).}
      \label{fig:summary}
\end{figure}

This procedure permits to obtain a value for $\kappa_{\rm dep}$. Finally $\kappa_{\rm wash}$ can be obtained by solving the Boltzmann equations in a way similar to the usual thermal leptogenesis, but using the reheating temperature $T_{\rm reh}$ as an initial temperature. Iterating this procedure over $\alpha_n$ and $\beta$, we obtain the plots in Fig.\,\ref{fig:summary}, in the case $M_N = 5\times 10^9$ GeV. For the PT sector, we consider either that the phase transition is catalized by a singlet scalar (left panel) coupling only to $\Phi$ or that $B-L$ is gauged and that the PT is catalized by the $B-L$ gauge bosons (right panel). The two results mostly differ because of the different depletion channels existing in those two scenario. The grey bands show the amount of enhancement we obtain compared to the conventional thermal leptogenesis, and the horizontal axis shows the strength of the supercooling, $\alpha_n$.
We obtain a $\mathcal{O}(20)$ enhancement compared to the thermal scenarios. 

The fraction of RHNs entering into the bubble, $\kappa_{\rm pen}$ is order one in the whole parameter space, but slightly decreases when $\alpha_n \sim 1$. On the other hand, for stronger phase transition and thus more drastic departure from equilibrium $\alpha_n >5$, $\kappa_{\rm wash}\simeq 1$. In this regime, the washout suppression inherent to thermal leptogenesis is avoided. 
We find that $\kappa_{\rm dep}$ causes a stronger suppression compared to $\kappa_{\rm wash}$, highlighting the importance of including these new annihilation channels, but it also becomes suppressed for larger $\alpha_n$. However, even if, at large values of $\alpha_n$, the washout and depletion effects become negligible, the dilution factor due to entropy injection, $(T_{\rm nuc}/T_{\rm reh})^3 = (1+\alpha_n)^{-3/4}$ strongly suppress the baryon yield.  
The final asymmetry is proportional to the all these factors and we find the enhancement is maximized around $\alpha_n \sim 5$ and $\beta \sim 60$.

We emphasize that such strong and slow FOPT would produce copious background of GW, that might be detectable at the ET observer for the model presented above.

\section{Leptogenesis via the production of heavy states}
\label{sec:leptogenesis_via_production}
In the former scenario, bubbles were critical for giving a sudden large mass to the RHN and thus suppressing the wash-outs due to inverse decays. The main novelty with respect to thermal leptogenesis was that the RHN received a Majorana mass from a FOPT. We now turn to more exotic mechanism where the bubble walls create very heavy states from the light states of in plasma\cite{Vanvlasselaer:2020niz}. 
\subsection{Production of heavy states with a fast bubbel wall}
We start by reviewing the process of heavy states production from fast expanding bubbles presented in\cite{Vanvlasselaer:2020niz}.   Let us assume the following Lagrangian
\bea 
\mathcal{L} =
|\d_\mu \Phi|^2+i \bar \chi \sl\d \chi + i \bar N \sl\d N- M_N \bar N N- Y \Phi \bar N \chi
\label{eq:Toy_model_1}
\eea
where $\chi$ a light
fermion and $N$ a heavy Dirac fermion with mass $M_N\gg \vev{\Phi}$, $M_N\gg T_{\text{nuc}}$ and $Y$ is the coupling between the scalar and the two fermions.  We work in the basis where fermion masses are real. In this setting, the equilibrium abundance of $N$ is exponentially suppressed. However in the case of an ultra-relativistic bubble expansion, the probability that the light $\chi$ fluctuates via mixing to the heavy $N$ is non-vanishing \cite{Vanvlasselaer:2020niz} and is approximately equal to
 \bea
 \mathcal{P}^{\rm tree}(\chi \to N) \approx  \frac{Y_{}^2 \vev{\Phi}^2}{M_N^2}\Theta(\gamma_w T_\text{nuc} - M_N^2 L_w)
 \label{eq:prod_tree}
 \eea
with $L_w \sim 1/m_\Phi$ the
thickness of the wall.
Thus, when the ultra-relativistic wall 
hits the plasma, it produces $N$ and $N^c$. Note that this abundance will be much larger than its equilibrium value. Indeed, outside of the bubble we have
\bea
&&n_{N_I}^{\vev{\Phi}=0}(T_{\text{nuc}}) \simeq  0, \hbox{~~~~Heavy $N_I$ have decayed}
\eea
while inside it became
\bea
\label{eq:n_prod}
n_{N_I}^{\vev{\Phi}\neq 0}&\simeq &
 \frac{1}{\gamma_w v_w}  \int \frac{d^3p}{(2\pi)^3} P_{\chi \to N}(p) \times f_\chi (p, T_{\text{nuc}}) 
\nn
& \simeq &
 \sum_i\frac{|Y_{iI}|^2 \vev{\Phi}^2}{ M_{I}^2 \gamma_w v_w}  \int \frac{d^3p}{(2\pi)^3}  \times f^{eq}_\chi (p,T_\text{nuc})\Theta ( p_z- M_{I}^2/\vev{\Phi})\nn
 &\simeq & \sum _i\theta_{iI} ^2 n_{\chi^i}^{\vev{\phi}=0}(T_{\text{nuc}}).
\eea
where we used all through the computation $v_w = \sqrt{1-1/\gamma_w^2} \approx 1 - \frac{1}{2\gamma_w^2}$ is the velocity of the wall.
We have defined the effective \emph{mixing angle}
\bea 
\theta_{iI}\equiv \frac{|Y_{iI}| \vev{\Phi}}{M_I}.
\eea

We now need to show that, at one loop level, interference within the bubble wall can create a difference abundance of $N$ and $ N^c$.

\subsection{CP violation in production}
To introduce CP violation in our production process, we will need to generalise the former Lagrangian of Eq. \eqref{eq:Toy_model_1} to include several families of light species $\chi$ and heavy species $N$,
\bea 
\label{Eq:example-mod}
\mathcal{L} =
i \bar \chi_i P_R  \sl\d \chi_i + i \bar N_I \sl\d N_I- M_I \bar N_I N_I- Y_{iI} \phi \bar N_I P_R\chi_i  - y_{I\alpha}(H\bar L_\alpha ) P_R N_I + h.c,
\eea
where $H$ and $L_\alpha$ have already been defined in section \ref{sec:bubble_assisted}, $P_R, P_L$ are the  chiral projectors, that we now make explicit. 
We choose this assignment of chirality in agreement with our further toy models.
Notice the difference with the Lagrangian in \eqref{Eq:L_RHN} where $\Phi$ was giving a mass to the RHN, while here we assume that another mechanism provides a mass to the RHN. In this sense, the transition of the $\Phi$ in the present scenario occurs \emph{after} the usual leptogenesis from the decay of heavy $N$.

\subsubsection{Calculation of the light to heavy transition \emph{at 1-loop level}}
\label{sec:asymmetry}
Let us now compute the asymmetries in the populations of the various particle immediately after the 
PT in the case of the model in Eq.\eqref{Eq:example-mod}. 
\begin{figure}
    \centering
    \includegraphics{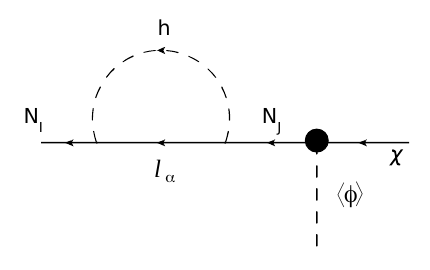}
    \caption{The diagram contributing to the function $f^{(HL)}$. }
    \label{fig:diaglepto}
\end{figure}

We first notice that, for the same reason than in usual out-of-equilibrium decay, CP violation cannot appear at tree level since processes will be proportional to $|Y_{iI}|^2$. As a consequence, we need to consider one loop corrections to it. To simplify the computations, we specialize to the case of ultra-relativistic bubble walls. First of all we need to know the CP violating effects in the $\chi_i\to N_I$
transition. Such effects appear due to the interference between tree-level and loop level diagram and scale like 
\bea
&&A(\chi_i\to N_I)_{\rm tree}\propto Y_{iI}\nn
&&A(\chi_i\to N_I)_{\rm 1-loop}\propto\sum_{k,J} Y_{iJ}Y_{k J}^*Y_{k I}\times f^{(\chi \phi)}_{IJ}+\sum_{\alpha,J} Y_{iJ}y_{\alpha J}^*y_{\alpha I}\times f^{(HL)}_{IJ}
\eea
where the functions $f^{(HL)}$ and $f^{(\chi\phi)}$ refer to the loop diagrams with virtual $\chi,\phi$ and $HL$ respectively. The computation of the loop shown in Fig.\eqref{fig:diaglepto} in the background of the bubble wall gives
\bea
 && \epsilon_{Ii} \equiv  \frac{| \mathcal{M}_{i \to I}|^2 -| \mathcal{M}_{\bar i\to \bar I}|^2}{\sum_i | \mathcal{M}_{i \to I}|^2 +| \mathcal{M}_{\bar i \to \bar I}|^2} \nn\\
 &&=\frac{2\sum_{k,J} {\rm Im} (Y_{iI} Y_{iJ}^*Y_{k J}Y^*_{k I} ) {\rm Im }f^{(\chi\phi)}_{IJ}}{ \sum_{i}|Y_{i I}|^2}+\frac{2\sum_{\alpha,J} {\rm Im} (Y_{iI} Y_{iJ}^*y_{\alpha J}y^*_{\alpha I} ) {\rm Im}f^{(HL)}_{IJ}}{ \sum_{i}|Y_{i I}|^2},
 \label{eq:CP_asym_1}
\eea
where the loop functions take a form reminiscent from the out-of-equilibrium decay ones: 
\begin{align}
\label{Loop-functions}
& \text{Im}[f^{(HL)}_{IJ}(x)] =\frac{1}{16\pi} \frac{\sqrt{x}}{1-x}, \qquad x = \frac{M_J^2}{M_I^2}
\\
& \text{Im}[f^{(\chi \phi)}_{IJ}(x)] = \frac{1}{32\pi} \frac{1}{1-x}.
\end{align}

Summing over the
flavours of $\chi_i$ we arrive at the following asymmetry in $N_I$ abundance
\bea
\epsilon_I\equiv \sum_i \epsilon_{Ii}=
\frac{2\sum_{\alpha,J,i} {\rm Im} (Y_{iI} Y_{iJ}^*y_{\alpha J}y^*_{\alpha I} ) {\rm Im}f^{(HL)}_{IJ}}{ \sum_{i}|Y_{i I}|^2}.
\label{eq:CP_asym_2}
\eea

We notice that the contribution from the $\chi, \phi$ loop vanished after summing over the contributions. This shows that the passage of the bubble wall can  create a difference in the abundances of $N_I$ and $\bar N_I$ inside the bubble. 

However since the $N_I$ are produced by $1\to 1$  transitions (conserving the total number of particles) the same difference will be present inside the bubble also for the abundances of $\bar \chi_i$ and $\chi_i$, which means that some of the abundance of $\chi_i$ has been removed from the plasma:
\bea
\sum_I \Delta n_{N^I} = - \sum_i \Delta n_{\chi^i} \Rightarrow  \qquad \sum_I \l(\Delta n_{N^I}-\Delta n_{\bar N^I} \r)= - \sum_i \l(\Delta n_{\chi^i}-\Delta n_{\bar\chi^i}\r) ,
\eea
where $\Delta n_{N,\chi}$ are the differences in abundances of the particles in the broken and unbroken phases. The passage of the wall still did not create $L$ number but separated it in a heavy and a light sector.

\begin{figure}
 \centering
 \includegraphics[scale=0.25]{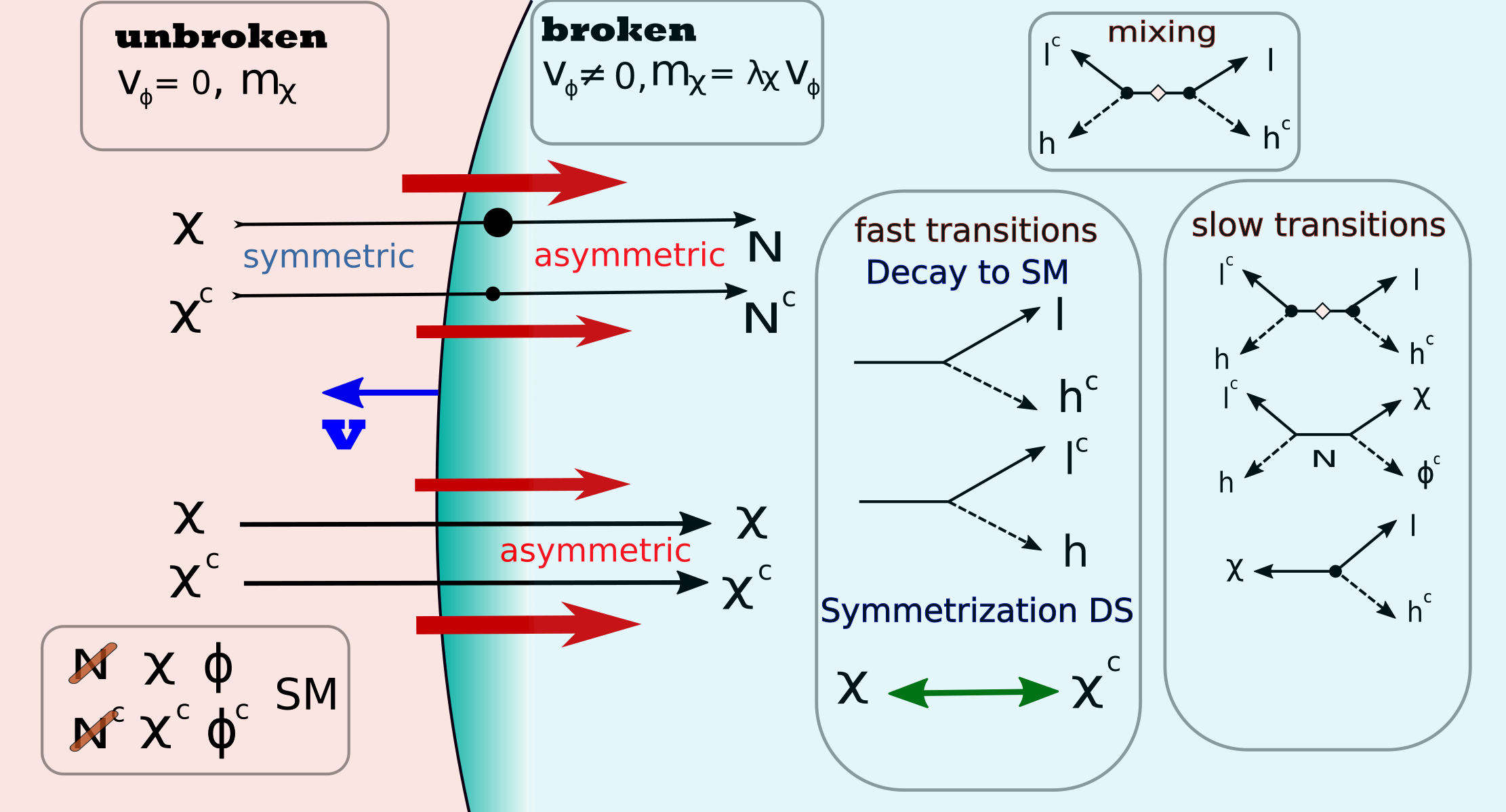}
 \caption{Mechanism at play in the phase transition-induced leptogenesis. Picture from \cite{Azatov:2021irb}. }
 \label{fig:diagram_1}
 \end{figure}

We can now move to the full model of leptogenesis with a fast bubble wall. Let us consider the following extension of the Lagrangian in Eq.\ref{Eq:example-mod}, where we have introduced $\Phi$ -dependent Majorana mass for the field $\chi$ and kept the rest of the interactions the same. 
Interestingly, only one specie of the Majorana fermion $\chi$ is sufficient for the generation of CP phase:
\bea
\label{Eq:ToyMod}
\mathcal{L}_{\text{int}}& = & \underbrace{\sum_{I} \bigg(Y_{I}(\phi^\dagger \bar{\chi}) P_L N_{I} + Y_{I}^{\star}\bar{N}_{I}P_R(\phi \chi)   \bigg) - V(\phi) + \frac{1}{2}\lambda_\chi \phi \bar{\chi}^c \chi + \sum_{I} M_I \bar N_I N_I}_{\text{Toy model of Dark Sector}} 
\\ \nonumber
&+ &\underbrace{\sum_{\alpha I} y_{\alpha I} (h \bar{l}_{\alpha,SM})P_R N_{ I} + h.c.}_{\text{Connection to SM}},
\eea 
We give the following $U(1)$ assignments, $L(\chi) = -1, L(N) = 1$ and $L(\phi)=2$, which respect $U(1)$ lepton number.  This $U(1)$ symmetry is obviously broken after the phase transition. 
The generation of the baryon asymmetry is illustrated in Fig.\ref{fig:diagram_1} and proceeds as follows: The expansion of the bubble generates an asymmetry in $N$ and $\chi$, with an opposite sign. Immediately after the transition, the asymmetry in $\chi$ is washed out due to the lepton-number violating 
Majorana mass term. The asymmetry is then transmitted to the SM when the $N$ decay via $N\to HL $,
and produce
\bea
 \label{eq:asym_model}
 \frac{n_{L}-n_{L^c}}{s}  &\simeq & \frac{1}{s(T_{\rm reh})}
\sum_{iI}\epsilon_{Ii} \frac{3\zeta(3) |Y_{iI}|^2 T_\text{nuc}^3 \vev{\phi}^2}{4\pi^2M_I^2} \times 
\frac{Br(N_I \to HL)}{Br(N_I \to HL)+Br(N_I \to \chi \Phi)}
\\ \nonumber
 &\simeq &
 \frac{ 135 \zeta(3) g_\chi}{8 \pi^4 g_\star}\bigg(\frac{T_{\rm nuc}}{T_{\rm reh}}\bigg)^3 \sum_I \theta^2_I
 \frac{2\sum_{\alpha,J} {\rm Im} (Y_{I} Y_{J}^*y_{\alpha J}y^*_{\alpha I} ) {\rm Im}f^{(HL)}_{IJ}}{ |Y_{ I}|^2} 
{\frac{\sum_\alpha | y_{\alpha I}|^2}{\sum_\alpha |y_{\alpha I}|^2+{ | Y_{I}|^2}}} \,, 
 \eea
where $g_*$ is total number of degrees of freedom and $s(T) = \frac{2\pi^2}{45}g_\star T^3$, and $g_\chi$ is the number of degrees of freedom of $\chi$ particle. 
 On the top of the CP violation in the $N$ production, there is also a CP violation in the decay of $N$,  
\bea
\l.\frac{n_{L}-n_{L^c}}{s}\r|_{\rm decay}\sim 
\sum_I\frac{\theta^2_I}{g_\star} \epsilon_{\rm decay}^I \bigg(\frac{T_{\rm nuc}}{T_{\rm reh}}\bigg)^3\times 
{\frac{\sum_\alpha | y_{\alpha I}|^2}{\sum_\alpha |y_{\alpha I}|^2+{ | Y_{I}|^2}}} \,. 
\eea
This constitutes a second source of asymmetry for the system, via the usual CP-violating decay. We will see that the dominant contribution depends on the different couplings of the systems.
This asymmetry in return is passed to the 
baryons by sphalerons, similarly to the original leptogenesis models \cite{Fukugita:1986hr}. And adding the contribution from the production and from the decay, we obtain 
\bea
\frac{\Delta n_B}{s} \equiv \frac{n_{B}-n_{\bar B}}{s} \simeq&& -\frac{28}{79}\times \frac{135 \zeta(3) g_\chi}{8\pi^4 g_*}\times \sum_I{\theta^2_I} \sum_{\alpha,J} {\rm Im} (Y_{I} Y_{J}^*y_{\alpha J}y^*_{\alpha I} ) {\rm Im}f^{(HL)}_{IJ}\nn
&&\times \l( \frac{2}{|Y_{ I}|^2}-\frac{1}{\sum_{\alpha} |y_{\alpha I}|^2}\r)\bigg(\frac{T_{\rm nuc}}{T_{\rm reh}}\bigg)^3
{\frac{\sum_\alpha | y_{\alpha I}|^2}{\sum_\alpha |y_{\alpha I}|^2+{ | Y_{I}|^2}}} \,. 
\label{eq:asym_model_1}
 \eea
The prefactor $-\frac{28}{79}$  comes 
from the sphalerons rates (see \cite{Harvey:1990qw}).
By assuming $O(1)$ parameters in the potential and dominant latent heat, $T_{\rm reh}\sim v.$  
 The factor ${\frac{\sum_\alpha | y_{\alpha I}|^2}{\sum_\alpha |y_{\alpha I}|^2+{| Y_{I}|^2}}}$ appears since a part of the asymmetry in $N$ is decaying back to $\phi \chi$.

Let us examine various bounds on the construction proposed. 
The non-vanishing VEV in the Lagrangian \eqref{Eq:ToyMod} generates a dimension 5 Weinberg operator of the see-saw form 
\cite{Minkowski:1977sc,Yanagida:1979as,GellMann:1980vs,Glashow:1979nm,Mohapatra:1979ia}
\bea 
\mathcal{O}_W = \sum_{I ,\alpha, \beta}\theta_{I}^2\frac{y_{\alpha I}y^*_{\beta I}(\bar{L}^c_\alpha H)(L_\beta H )}{m_\chi}
\eea
which induces a mass for the heaviest light neutrinos
\bea
\Rightarrow\text{Max}[m_{\nu}] \sim \text{Max}\bigg[\sum_I|y_{\alpha I}|^2\theta^2_{I}\bigg] \frac{  v_{EW}^2 }{m_\chi}.
\label{eq:neutrino_mass}
\eea
Combining Eqs. \eqref{eq:asym_model_1}, 
\eqref{eq:neutrino_mass} with 
observed neutrino mass scale and the constraints
$\text{Max}[\theta^2_{I}] \gtrsim 10^{-5}, y \sim \mathcal{O}(1)$, we obtain the following constraints
\bea 
\Rightarrow \qquad  m_\chi \gtrsim 5\times 10^{9} \text{GeV}  \qquad \Rightarrow \qquad  \vev{\Phi} \gtrsim 10^{9} \text{GeV}.
\label{constraints_1}
\eea
The inverse decay due to collisions $HL \to N$ will efficiently erase the asymmetry. 
The Boltzmann equation controlling this wash out is 
\bea
\label{eq:condition2to1}
\frac{d Y_{\Delta_\alpha}}{d z}\simeq -\frac{0.42 e^{-z}z^{5/2}}{g_*^{1/2}g_\alpha}\l(\frac{M_p}{m_\chi}\r)\l(\frac{g_\chi \Gamma_\alpha}{m_\chi}\r)Y_{\Delta_\alpha}, \qquad \Gamma_{\alpha} \approx \l|\sum_{I} y_{\alpha I}\theta_{I} 
\r|^2\frac{m_\chi}{8\pi g_\chi}.
\eea
from which we obtain that $Y_{\Delta_\alpha}$ remains invariant for $m_\chi/T_{\rm reh}\gtrsim 15$ (for the scale $m_\chi\sim 10^{9}$ GeV). The following approximate relation for the minimal $m_\chi/T_{\rm reh}$ to avoid wash out is valid
\bea 
\frac{m_\chi}{T_{\rm reh}}  \gtrsim \log \frac{M_p}{m_\chi} 
-9
\label{eq:hierarchy}
\eea
where we took $\theta_I \sim 10^{-2}$ as a typical value.
The $\Delta L = 2$ lepton violating operator $LLHH$, if it enters in equilibrium will also erase the initial asymmetry
\begin{align}
\Gamma(H^c L_\alpha \to HL_\beta^c )(T) \approx \frac{4}{1.2 \pi^2g_\alpha }\sum_{ i I }\frac{\theta_{iI}^4}{m_\chi^2}y^2_{i\alpha}y^2_{i\beta} T^3
\approx \frac{2}{1.2\pi^2 }\l(\frac{m_\nu}{v_{EW}^2}\r)^2T^3 \, ,
 \nonumber \\ 
\Rightarrow 
T_{\rm reh}\lesssim 5\sqrt{g_\star}\frac{v_{EW}^4}{M_p m_\nu^2}\sim 5\times 10^{12
} \text{ GeV}\, ,
\end{align}
where we took $m_\nu^2 \sim 0.0025 \text{ eV}^2$.

In conclusion we can see that this 
construction can lead to a viable mechanism of
leptogenesis if there is a mild 
hierarchy between the scales; $M_I > \vev{\Phi}$ and $m_\chi,M_I> T_{\rm reh}$.  Finally, we obtain that the matching with the light neutrino masses makes 
this mechanism  operative in the range of scales $10^9 \text{ GeV} < \vev{\Phi} < 5\times 10^{12}$ GeV.

\section{EW baryogenesis via PT with fast bubble walls}
\label{sec:model2}
\begin{figure}
 \centering
 \includegraphics[scale=0.25]{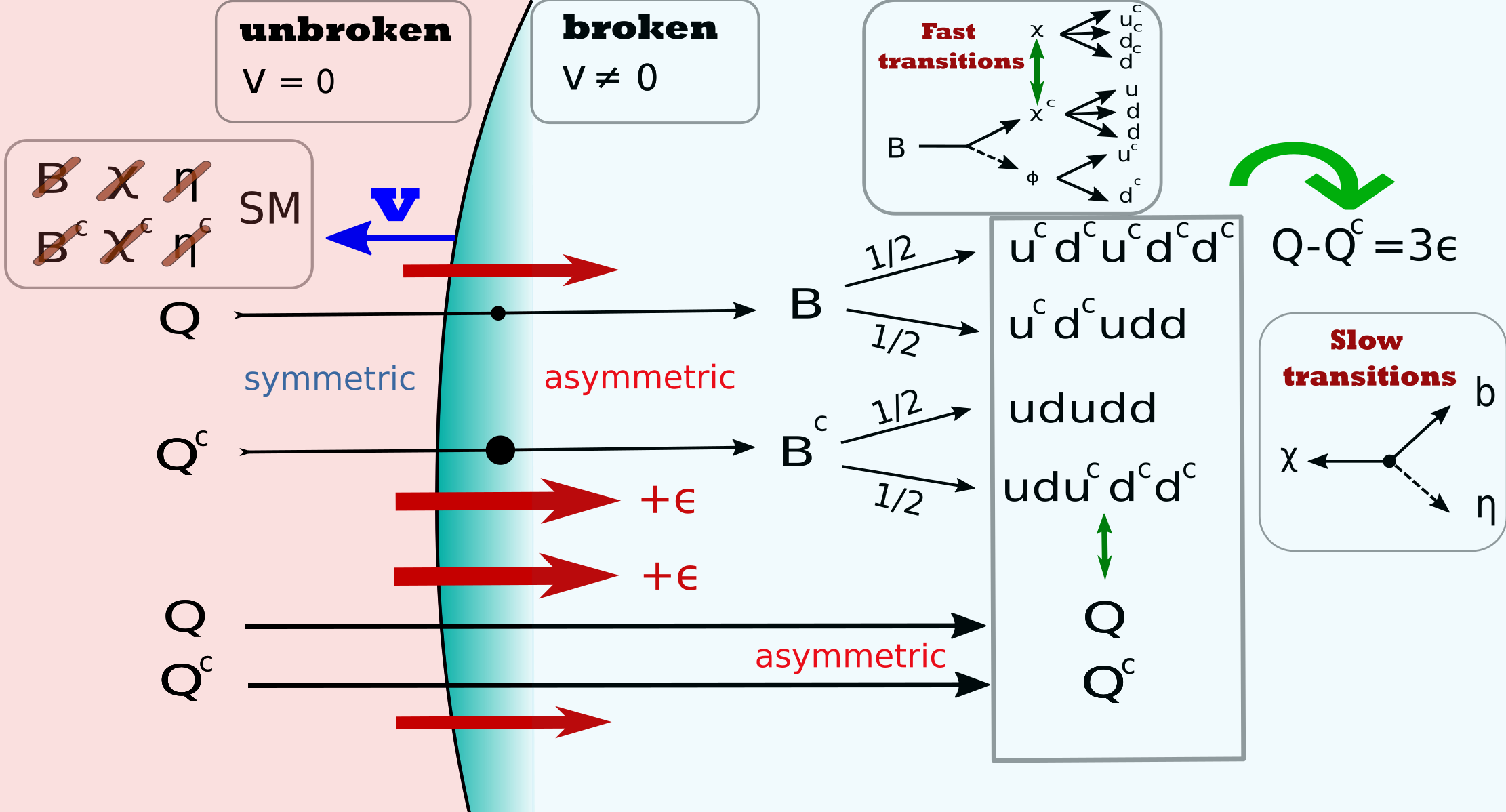}
 \caption{Mechanism at play in the low energy baryogenesis, from \cite{Azatov:2021ifm}.   }
 \label{fig:diagram_2}
 \end{figure}

In the previous sections, we have discussed the high energy leptogenesis catalized by the passage of a fast bubble wall. ElectroWeak baryogenesis also relies on the out-of-equilibrium situation surrounding the bubble wall, but it is efficient \emph{only if the velocity of the bubble wall is not much faster than the sound speed}\cite{Laurent:2020gpg}. On the other hand, gravitational wave amplitude is typically maximized for large velocities $v_w \to 1$\cite{Caprini:2019egz}, which makes large GW signal and efficient baryogenesis more or less mutually exclusive. 
 
 However in this section, we would like to use the mechanism proposed above, using large velocities and thus predicting copious amount of GW, to build a model of EWBG. One robust prediction of such a scenario is the large amount of GW emitted at the transition, with peak frequency fixed by the scale of the transition $f_{\text{peak}}\sim 10^{-3}\frac{T_{\text{reh}}}{ \text{GeV}}$ mHz (see \cite{Weir:2017wfa} for review). Such SGWB signal could be detected in future GW detectors such as LISA\cite{amaroseoane2017laser,Caprini:2019egz}, eLISA\cite{Caprini:2015zlo}, LIGO\cite{vonHarling:2019gme, Brdar:2018num}, BBO\cite{Corbin:2005ny, Crowder:2005nr}, DECIGO\cite{Seto:2001qf, Yagi:2011wg, Isoyama:2018rjb}, ET\cite{Hild:2010id,Sathyaprakash:2012jk,Maggiore:2019uih}, AION\cite{Badurina:2019hst}, AEDGE\cite{Bertoldi:2019tck}. This array of observers will be able to probe GW with frequencies in the window of mHz to kHz, which is the optimal scale for this mechanism to take place.

 Below we present a prototype model, which we can consider as a first simple toy model:
\bea
\label{eq:modelB}
{\cal L} =&&{\cal L}_{SM} +  m_\eta^2 |\eta |^2+\sum_{I=1,2} M_{I} \bar B_I B_I \nn
+&&\l(\sum_{I=1,2}Y_I (\bar B_I H)P_L Q+
{y_I \eta^*\bar B_I P_R\chi}
+ \kappa \eta^c d u
 +\frac{1}{2} m_\chi \bar{\chi^c} \chi + h.c. \r) .
\eea
 
The model contains a Majorana field $\chi$ and two  vector-like $B$ quarks with the masses $M_{1,2}\sim m_\chi$.
 $\eta$ is a scalar field (a diquark) which is in the fundamental representation of QCD with electric charge $Q(\eta)=1/3$, $Q,u,d$  are  the  SM quark doublet and singlets respectively, we ignore the flavour indices for now, 
We assume that the EW phase transition is of the first order  and that the bubble wall becomes relativistic, we will discuss later how to build such scenario. For the reasons that we will explain later, we need to assume that only the third generations couples to the heavy vector like $B$ quark. 
Notice however that the interaction  $ L H \chi^c$ allowed by the gauge symmetries of the model but would violate the baryon number of one unit and lead to proton decay. We set it to zero in order to avoid this as this feature can be attributed to some accidental discrete symmetry. 
The baryon number are as follows: $B(\eta)=2/3, B(\chi)=1$, so that the $m_\chi$ violates the baryon symmetry by two units. The sweeping of the relativistic wall, via the collision of the b-quarks with bubbles, produces $B_I, B_I^c$ and inside the bubble, we obtain the following abundances
\bea
n_{B_I}-n_{B^c_{I}}=-\theta_{I}^2 \epsilon_I n_b^0 \qquad \qquad 
n_{b}-n_{b^c}=\sum_I\theta_{I}^2 \epsilon_I n_b^0
\label{eq:asymmB}
\eea
where $n_b$ is the number density of the bottom-type quark, $\theta_I \approx \frac{Y_I \vev{H}}{M_I}$ is the mixing angle  and  $\epsilon_I$ is defined like in  Eq.
\eqref{eq:CP_asym_1} (in this case there is no $i$ index since we coupled it only to  the third generation of quarks). The imaginary part of the loop function generated by a diagram similar to the one of Fig.\ref{fig:diaglepto} becomes
\bea 
{\text{Im}[f^{IJ}_B(x)]} =
\frac{1}{32\pi}
\frac{M_I M_J}{M_I^2-M_J^2}\frac{\sqrt{(M_I^2 - m_\eta^2 + m_\chi^2)^2 - 4m_\chi^2M_I^2}}{M_I^4} 
\l( M_I^2+m_\chi^2-m_\eta^2\r)  \,. 
\label{eq:loop_func}
\eea 
Like in the leptogenesis scenario, the asymmetry is separated in the heavy ($B$) and light $b$ sector, with an opposite sign:
\bea
\label{eq:refBb}
\sum_I  \l(n_{B_I}-n_{ B_I^c}\r)
= -( n_{b}- n_{b^c}) .
\eea
Let us see what will happen after $B_I$ decays.
If the mass spectrum satisfies $M_I> m_\chi >M_\eta$, there are four different channels, two leading to wash-outs and two enhancing the asymmetries:  
 \bea
 &&(i)~~ \text{wash-out}:~~B_I\to \chi d^c u^c \to (b d u  d^c u^c) \qquad ~~B_I^c\to \chi^c d u \to (b^c d^c u^c  d u)\nn
 &&(ii)~~\text{asymm. generation}:~~B_I\to \chi^c d^c u^c \to (b^c d^c u^c  d^c u^c) \qquad ~~B_I^c\to \chi d u \to (b d u  d u).\nn
 \label{eq:cascade}
 \eea
 As a result the asymmetry between SM quarks and antiquarks will be given 
  \begin{align} 
(n_{q}-n_{q^c}) &=\sum_I (n_{B_I}-n_{ B_I^c}) 
 \l[\l(-\frac{5}{2}+\frac{1}{2}\r)Br(B_I\to \chi \eta^c)+Br (B_I\to b h) 
\r]+(n_b-n_{b^c})\nn,
 &=-3\sum_I (n_{B_I}-n_{ B_I^c}) Br(B_I\to \chi \eta^c),
 \end{align}
where we have used $Br(B_I\to \chi \eta^c)+Br (B_I\to b h) =1$ and Eq.\ref{eq:refBb} to derive the last relation.
Finally, for the total baryon asymmetry we obtain
 \bea
 \frac{\Delta n_{Baryon}}{s}  
\approx&&\frac{135 \zeta(3)}{8 \pi^4} \sum_{I,J}\theta_I^2   \frac{|y_I|^2}{|y_I|^2+|Y_I|^2}\times \frac{g_b}{ g_\star}\bigg(\frac{T_{\rm nuc}}{T_{\rm reh}}\bigg)^3\nn
&&
\times {\rm Im}  (Y_I Y^*_J y_I^*y_J )
\l(-\frac{2 {\rm Im} [f_B^{IJ}]}{|Y_I|^2}+\frac{4 
{\rm Im} [f_B^{IJ}]|_{m_{\chi,\eta}\to 0}
}{|y_I|^2}
\r).
 \eea
To match the observed baryon abundance, we need that
\bea 
\Rightarrow \boxed{\theta_I^2 \bigg(\frac{T_{\rm nuc}}{T_{\rm reh}}\bigg)^3 \sim 10^{-(6-7)}}.
\eea

After this phase of fast decay, slow transition mediated by the heavy states can still wash out the asymmetry. They can be of two types:
 \bit
 \item $b\eta \to \chi $:  The Boltzmann equation controlling the interaction $b\eta \to \chi $ is
\bea
&&\frac{d \epsilon_q}{d z}=-
\frac{0.42 e^{-z}z^{5/2}}{g_*^{1/2}g_q}\l(\frac{M_p}{m_\chi}\r)\l(\frac{g_\chi \Gamma(\chi\to \eta b)}{m_\chi}\r)\epsilon_q
\nn
&&\Gamma(\chi\to \eta b) \approx \l|\sum_{I} y_I \theta_I\r|^2\frac{m_\chi(1-m_\eta^2/m_\chi^2)}{8\pi g_\chi} \label{washout}
\eea
The requirement that this process is decoupled imposes a constraint on the mass of the field $\chi$: $m_\chi/T_{\rm reh}\gtrsim 30$. At the end of the day, the main constraint on the mass spectrum of our model is
\bea
\boxed{\frac{m_{B,\chi,\eta}}{T_{\rm reh}}\gtrsim 30.}
\eea
This translates to bound $m_{B,\chi,\eta} \gtrsim 3$ TeV, or to a wall boost factor of the form 
\bea 
\frac{m_{B,\chi,\eta}^2}{T_{\rm nuc}v } \gtrsim \gamma_w
\eea 
or 
\bea
\boxed{\gamma_w \gtrsim 10^3 \, . }
\eea

Using Eq.\eqref{eq:terminal_velo}, we observe that this corresponds to a supercooling 
\bea 
T_{\rm nuc} \lesssim 10 \text{ GeV} \, . 
\eea 

\item $ddu  \leftrightarrow d^c d^c u^c $: Integrating out all the new heavy fields $B, \chi, \eta$ also generates new dangerous operators of the form
\bea
\frac{ddu \overline{d^c d^c u^c}}{M_\eta^4}\times \frac{1}{m_\chi}\times \theta^2
\qquad \Rightarrow \qquad
{\frac{1}{4\pi^5}\bigg(\frac{1}{16\pi^2}\bigg)^2}\frac{T_{\rm reh}^{11} }{M_\eta^8 m_\chi^2}\theta^4\lesssim \frac{T_{\rm reh}^2}{M_{p}}.
\eea
However, this interaction, in all our parameter space, is always decoupled and do not bring any further wash out. 
\eit

This low-energy model has the interesting consequence that it induces potential low-energy signatures. In this section, we enumerate those possible signatures without assuming that $Q, u,\text{and } d$ are the third generation quarks.  
\paragraph{ $n-\bar n$ oscillations}
Integrating out the heavy states, we obtain the following operator\cite{Fridell:2021gag}.
 \bea 
\frac{1}{\Lambda^5_{n\bar{n}}}\overline{u^c d^cd^c} udd  \equiv\frac{(\sum{\kappa \theta_I y_I})^2}{M_\eta^4 m_\chi }\overline{u^c d^cd^c} udd  \quad \Rightarrow \quad \delta m_{\bar{n}-n} \sim \frac{\Lambda_{QCD}^6}{M_\eta^4 m_\chi}(\sum{\kappa \theta_I y_I})^2.
\eea

 Current bounds on this mixing mass are of order $ \delta m_{\bar{n}-n}  \lesssim 10^{-33} $ GeV~\cite{BaldoCeolin:1994jz, Abe:2011ky,Rao:1982gt, Buchoff:2012bm,Syritsyn:2016ijx}.
It is very restrictive if $B$ couples to light quarks, we conclude that we need to require that $B$ only couples to the third family. Depending on the flavor of $Q,u,d$ (the first or the third family) our scenario can be tested in the future experiment~\cite{Phillips:2014fgb,Milstead:2015toa, Frost:2016qzt, Hewes:2017xtr}. 

\paragraph{ Flavor violation}
 For the $\eta$-diquark field that we used in this model, the FCNC are absent at 
tree level\cite{Giudice:2011ak}. The loop level effects will however unavoidably lead to strong constraints if $\eta d u$ coupling contains the light generation quarks \cite{Giudice:2011ak}. FCNC also forces $B$ to couple mostly to the third family.

\paragraph{ Bounds from EDMs}
EDM are also typical signature of CP violation if it occurs at lowe energy {
 \bea
 -i \frac{g_3\tilde d_q}{2} \bar Q \sigma^{\mu \nu}T^A\gamma_5 QG^A_{\mu \nu}
 \eea
 which is the chromo-electric dipole moment (see \cite{Engel:2013lsa} for a review). 
   The dipole in our model can be estimated to be
 \bea 
 \frac{d_e}{e} \sim \frac{m_e (y Y e)^2}{(4\pi)^6}\bigg(\frac{1}{\Lambda_{EDM}^2}\bigg) \sim 3\times 10^{-33}\times \l(\frac{10 {\rm TeV}}{\Lambda_{EDM}}\r)^2 {\rm cm}
 \eea 
 which is much below the current experimental bound
 \cite{Andreev:2018ayy} $|d_e|< 1.1\times 10^{-29} {\rm cm} \cdot e$.

}

So far we have left the phase transition sector inducing a fast wall for the EWPT undefined. The necessary ingredient  for the mechanism is a strong first order electroweak phase transition and various studies 
indicate that even a singlet scalar
(see ref.\cite{PhysRevD.45.2685,Choi:1993cv,Espinosa:1993bs,Profumo:2007wc,Espinosa:2011ax,Chen:2017qcz,Ellis:2018mja}) 
extensions of SM can help making the EWPT strongly first order. In \cite{Azatov:2022tii}, authors studied a singlet augmented SM in the case of a two steps phase transition. We study the following two steps PT
\bea
(0,0)\to (0,\vev{s}) \to (v_{EW},0)\;,  
\eea
and focus on the region which induce relativistic bubbles. 
This pattern can occur if the $m_s^2$ parameter is positive. We will consider the following simple model
\bea
V_{\rm tree}(H, s)
= - \frac{m_h^2}{4}H^2 +\frac{m_H^2}{8v_{EW}^2}H^4  -\frac{m_s^2}{4}s^2 +\frac{\lambda_{hs}}{4} s^2 H^2 +\frac{m_s^2}{8v_s^2}s^4\;.
\label{eq:potential_1}
\eea
where $v_{EW}= \sqrt{m_H^2/2\lambda}$ GeV and $v_s=\sqrt{m_s^2/2\lambda_s}$ correspond to the local minima at $(\vev{ H}=v_{EW},{\vev{s}=0)}$ and  $(\vev{H}=0,{\vev{s}=v_s)}$ respectively. 
The origin of two-step PT can be intuitively understood from the 
following considerations. For simplicity let us ignore the Coleman-Weinberg potential and restrict the discussion  by considering only the thermal 
masses. Then the potential will be given by 
 \bea
 \label{eq:pot-app}
 V({{\cal H}, s})&\approx& V_{\rm tree}({\cal H}, s)+\frac{T^2}{24}\l[\sum_{bosons} n_i M_i^2({\cal H}, s))+\frac{1}{2} \sum_{fermions}n_F M_F^2({\cal H}, s)\r]\;,\nonumber\\
&=& V_{\rm tree}({\cal H}, s) + T^2\l[h^2\l(\frac{g'^2}{32}+\frac{3 g^2}{32}+\frac{m_h^2}{8 v_{EW}^2}+\frac{y_t^2}{8}+\frac{\lambda_{hs}}{48}\r)+s^2\l(\frac{m_s^2}{16 v_s^2}+\frac{\lambda_{hs}}{12}\r)\r].\nonumber\\
 \eea
 From this expression we can clearly see that the temperatures when the minima with non zero vevs appear for the Higgs and singlet fields can be different. Then it can happen that the $Z_2$ breaking phase transition occurs before the EW one. This means that there will be first a phase transition from $(0,0)\to (0, v_s)$. We can now focus on the second PT, which is the real EWPT, $(0,v_s)\to (v_{\rm EW}, 0)$. Due to a tuning of the term $ - \frac{m_h^2}{4}H^2$ against $\frac{\lambda_{hs}}{4} v_s^2 H^2$, the potential can become very flat around the false vacuum, leading to supercooling. 
 
\begin{figure}
\centering
  \includegraphics[scale=.4]{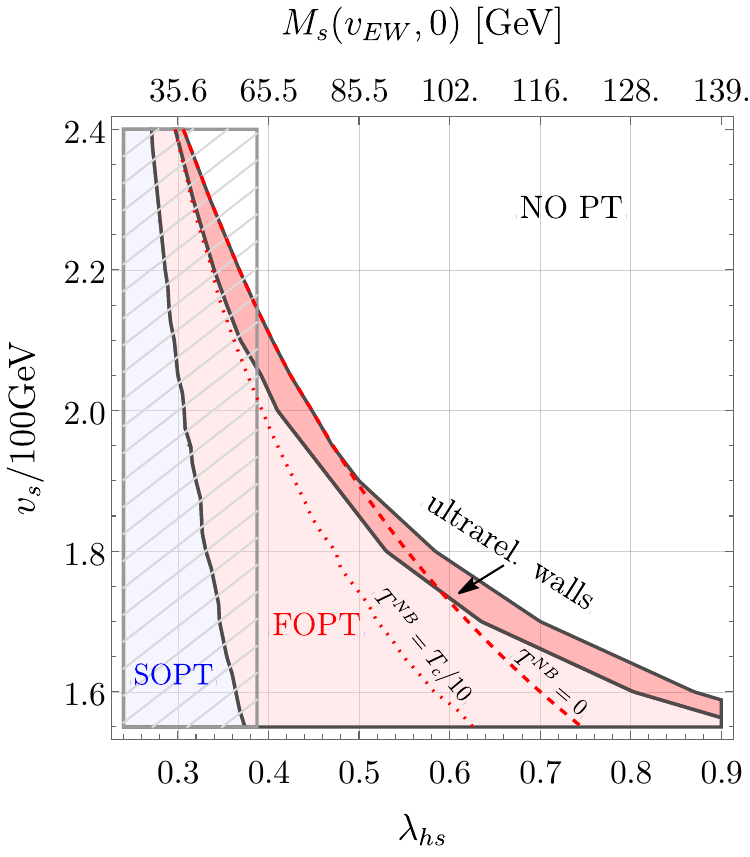}
  \includegraphics[scale=0.8]{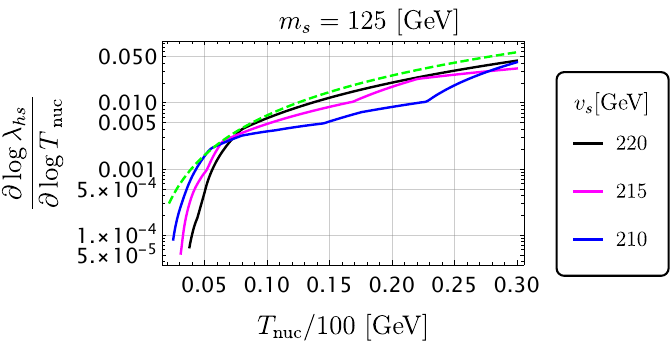}
 \caption{Left: scan of the parameter for the second PT.  Right: Tuning of the coupling $\lambda_{hs}$ as a function of the nucleation temperature. The dashed green line represents the naive tuning $\sim (T_{\rm nuc}/m_h)^2$.  }
  \label{fig:scan}
\end{figure}

 On Fig.\ref{fig:scan}, we show the scan of the parameter of the second PT (Left) and the tuning required to obtain a given $T_{\rm nuc}$ (Right). We observe that having $T_{\rm nuc} \sim 10 $ GeV and leading to $\gamma_w \sim 10^3$ requires only moderate $\mathcal{O}(0.01)$ tuning. 

An interesting follow-up would be to combine such scenario with the production of DM proposed in \cite{Azatov:2021ifm}.

\section*{Acknowledgements}
 MV is supported by the ``Excellence of Science - EOS" - be.h project n.30820817, and by the Strategic Research Program High-Energy Physics of the Vrije Universiteit Brussel.

\appendix

\bibliographystyle{JHEP}
{\footnotesize
\bibliography{biblio}}

\end{document}